\documentclass[journal,twocolumn]{IEEEtran}
\usepackage{epsfig,makeidx,color}
\usepackage{amsmath,amssymb,bbm}
\usepackage{cite,graphicx,lipsum}
\usepackage{enumerate}
\usepackage[switch,pagewise]{lineno}
\usepackage{hyperref}
\hypersetup{
        colorlinks = true,
        citecolor=blue,
}
\def\cK{{\cal K}}

\def\uP{{\mathbb P}}

\def\uE{{\mathbb E}}

\newtheorem{mytheorem}{\bf Theorem} % [section]
\newtheorem{mylemma}{\bf Lemma} % [section]
\def\be{ \begin{equation} }
\def\ee{ \end{equation} }
\def\bea{ \begin{eqnarray} }
\def\eea{ \end{eqnarray} }

\def\by{{\bf y}}
\def\bc{{\bf c}}

\def\bs{{\bf s}}
\def\ba{{\bf a}}

\def\bn{{\bf n}}

\def\bC{{\bf C}}

\def\bI{{\bf I}}

\def\bR{{\bf R}}

\def\b0{{\bf 0}}

\def\cC{{\cal C}}

\def\cN{{\cal N}}
\def\cS{{\cal S}}

\ifCLASSOPTIONonecolumn
  \newcommand{\figwidth}{0.55\columnwidth}
  
\else
  \newcommand{\figwidth}{0.80\columnwidth}
  
\fi

\begin{document}

\title{On Improving 
Throughput of Multichannel ALOHA using Preamble-based Exploration}

\author{Jinho Choi\\
\thanks{The author is with
the School of Information Technology,
Deakin University, Geelong, VIC 3220, Australia
(e-mail: jinho.choi@deakin.edu.au).
This research was supported
by the Australian Government through the Australian Research
Council's Discovery Projects funding scheme (DP200100391).}}

%\linenumbers
%Under independent Rayleigh fading,

\maketitle
\begin{abstract}
Machine-type communication (MTC)
has been extensively studied to provide connectivity
for devices and sensors in the Internet-of-Thing (IoT).
Thanks to the sparse activity,
random access, e.g., ALOHA,
is employed for MTC to lower signaling overhead.
In this paper, we 
propose to adopt exploration for multichannel 
ALOHA by transmitting preambles before transmitting data packets
in MTC,
and show that the maximum throughput
can be improved by a factor of $2 - e^{-1} \approx 1.632$,
In the proposed approach,
a base station (BS) needs to send the feedback information 
to active users to inform
the numbers of transmitted preambles in multiple channels,
which can be reliably estimated as in compressive random access.
A steady-state analysis is also performed with fast retrial,
which shows that the probability of packet collision
becomes lower and, as a result, the delay outage
probability is greatly reduced for a lightly loaded system.
Simulation results also confirm the results from analysis.
\end{abstract}


\ifCLASSOPTIONonecolumn
\baselineskip 28pt
\fi

\section{Introduction}

In order to support the connectivity of a large number
of devices and sensors for the Internet of Things (IoT),
machine-type communication (MTC)
has been considered in cellular systems
\cite{Taleb12} \cite{3GPP_MTC} \cite{3GPP_NBIoT}.
In fifth generation (5G) systems,
it is expected to have more standards for
MTC \cite{Condoluci15} \cite{Shar15} \cite{Bockelmann16}.
In general, in order to support a large number of devices
(in this paper, we assume that devices and users are 
interchangeable)
with sparse activity (i.e., only a fraction of them 
are active at a time), random access is widely considered 
as it can avoid high signaling overhead.
In particular,
most MTC schemes are based on (slotted) ALOHA \cite{Abramson70},
and ALOHA is extensively studied for MTC as in
\cite{Arouk14} \cite{Lin14} \cite{Chang15} \cite{Choi16}.

Since ALOHA plays a key role in MTC,
various approaches are considered for ALOHA 
in order to improve the performance 
in terms of throughput
(which may result in the increase of the number of devices
to be supported). In \cite{Casini07},
contention resolution repetition diversity (CRRD) 
is considered 
together with successive interference cancellation (SIC)
for a better throughput. 
The notion of coding is applied to CRRD in \cite{Liva11}
\cite{Paolini15}, which results in coded random access,
where it is shown that the throughput 
(in the average number of successfully transmitted packets per slot)
can approach 1.
In \cite{Amat18}, a finite-length analysis (as opposed to
asymptotic analysis) is carried out for coded random access.
Coded random access is also considered for massive 
MTC with decoding that is based on statistical channel 
knowledge only in \cite{Duchemin19}.

Furthermore, in \cite{Choi_JSAC} \cite{Choi18b},
the notion of non-orthogonal multiple access (NOMA)
\cite{Choi08} \cite{Ding_CM}
is applied to ALOHA so that multiple virtual access channels
in the power domain are available without any bandwidth expansion,
and it is shown that the throughput can be significantly
improved at the cost of high power budget at users.
Note that as in coded random access, SIC has to be used
at a base station (BS) to remove interfering signals.
In \cite{Seo18}, the performance of NOMA-based random access
is further analyzed.

In this paper, we consider a different approach than coded random access
\cite{Paolini15} and NOMA-based random access \cite{Choi_JSAC}.
The proposed approach uses exploration for multichannel ALOHA to improve
the performance, while it does not require SIC at a BS.
Thus, whenever SIC is not affordable at a BS or a receiver, the proposed
approach can be used to improve the throughput.
As in multi-armed bandit problems
\cite{Bianchi06} \cite{Cohen07},
exploration can help improve the performance of multichannel ALOHA.
For exploration, in the proposed approach,
each active user (i.e., a user
with packet to transmit) is to send a preamble
prior to packet transmission,
and a BS sends the feedback information 
to active users to inform
the numbers of transmitted preambles in multiple channels.
We show that the feedback information,
which is the outcome of exploration,
can improve the throughput of multichannel ALOHA
(as well as single-channel ALOHA).
In particular, in terms of the maximum throughput,
it is shown that the performance can be improved by a factor
of $2 -e^{-1} \approx 1.632$ thanks to exploration.

It is noteworthy that the exploration by sending preambles
becomes possible if the BS is able to estimate
the number of transmitted preambles in each channel.
Thanks to the notion of compressive random access 
\cite{Applebaum12} \cite{Wunder14} \cite{Schepker15} \cite{Choi17IoT}
\cite{Choi_CRA18},
the BS can estimate the number of 
transmitted preambles in each channel precisely.
In addition, the proposed approach
does not use SIC and CRRD, which makes it easy to implement.
In general, while SIC is promising to improve the performance
of random access as mentioned earlier,
the implementation of SIC is known to be difficult.
In particular, since SIC requires to reproduce the received signals of
individual transmitted signals, a precise channel
estimation is required with propagation delay estimation
as well as frequency offset estimation (in uplink channels,
each user has different frequency offset and propagation delay).
Thus, a small BS or gateway has difficult to implement SIC
due to limited hardware and computing resources.
We believe that the proposed approach would be suitable for
the case that low-cost small base stations or gateways
are used for MTC.

We also consider fast retrial and analyze 
the steady-state performance to see the impact of exploration
on delay performance with fast retrial.
It is shown that the delay outage probability
can be significantly reduced by exploration 
when the system is lightly loaded.

In summary, the main contributions\footnote{A conference
version of the paper is \cite{Choi20}, where the main
idea of the proposed approach is presented 
without steady-steady analysis.}
of the paper
are as follows: 
\emph{i)} a multichannel ALOHA scheme is proposed
with preamble-based exploration to improve the throughput;
\emph{ii)} performance analysis is carried out,
which shows that the maximum throughput of the proposed
is higher than that of conventional ALOHA by a factor of
$2 - e^{-1}$;
\emph{iii)} using a steady-state analysis, 
it is shown that the exploration 
can significantly reduce
the delay outage probability
when fast retrial is employed 
for a lightly loaded system.

The rest of the paper is organized as follows.
In Section~\ref{S:SM},
we present the motivation and system model
for the proposed approach. To see the performance,
we consider the maximum throughput analysis in 
Section~\ref{S:MT}.
A steady-state analysis is carried out with fast retrial
in Section~\ref{S:SS}. We discuss some implementation
issues in Section~\ref{S:Imp}
and present simulation results in Section~\ref{S:Sim}.
The paper is concluded with remarks in Section~\ref{S:Conc}.

{\it Notation}:
Matrices and vectors are denoted by upper- and lower-case
boldface letters, respectively.
$\uE[\cdot]$ denotes the statistical expectation.
$\cC \cN(\ba, \bR)$
represents the distribution of
circularly symmetric complex Gaussian (CSCG)
random vectors with mean vector $\ba$ and
covariance matrix $\bR$.

\section{Motivation and System Model}	\label{S:SM}

Throughout this paper, we only consider a slotted ALOHA
system consisting of one BS and multiple users for uplink
transmissions,
where the BS periodically transmits a beacon signal 
for synchronization.

\subsection{Motivation}

Consider a single-channel ALOHA system.
Let $T_{\rm d}$ denote the length of data packet.
Throughout the paper, 
it is assumed that the data packets of users
have the same length.
If an active user (with a data
packet to transmit) knows that there are other active
users, she may not transmit to avoid collision.
In order to see whether or not there are 
other active users, suppose that
each user has a unique preamble and
an active user transmits its preamble
sequence before data packet transmission, 
which can be seen as the {\it exploration}
to learn the environment.
Let $T_{\rm p}$ denote the length of preamble.
It is assumed that $T_{\rm p} < T_{\rm d}$ in general.
At the end of preamble transmission, 
we assume that the BS is able to detect
all the transmitted preamble sequences
and sends a feedback signal to inform
the number of the transmitted preamble sequences.
The length of feedback signal is denoted by $T_{\rm f}$.

An active user can make a decision whether or not
she transmits her data packet based on the feedback from the BS.
If there is only one preamble transmitted, 
the active user should send a data packet as there is no
other active user. However, if the number of 
transmitted preambles is larger than $1$, 
each active user may transmit 
a packet with a certain probability that might be
less than 1.
For example, in order to maximize the probability of successful
transmission, 
the access probability might be $\frac{1}{K}$,
where $K$ represents the number of active users
or transmitted preambles that is fed back from the BS.
%Note that since $K$ is known by the feedback information
%from the BS, each active user can set its access probability
%to $\frac{1}{K}$.
Therefore, for a given $K \ge 1$, the throughput,
which is the average number of transmitted packets without collisions,
becomes
\be
\eta_{\rm sa} (K) = \left( 1 - \frac{1}{K} \right)^{K-1} \ge e^{-1}.
	\label{EQ:ineq}
\ee
As $K \to \infty$, we can see that 
$\eta_{\rm sa}$ approaches $e^{-1}$.
On the other hand, if $K = 1$, $\eta_{\rm sa} = 1$.
From \eqref{EQ:ineq}, the average throughput can be shown to be
higher than $e^{-1}$ as follows:
\be
\uE[\eta_{\rm sa} (K)] \ge e^{-1}.
	\label{EQ:esa}
\ee

\begin{figure}[thb]
\begin{center}
\includegraphics[width=\figwidth]{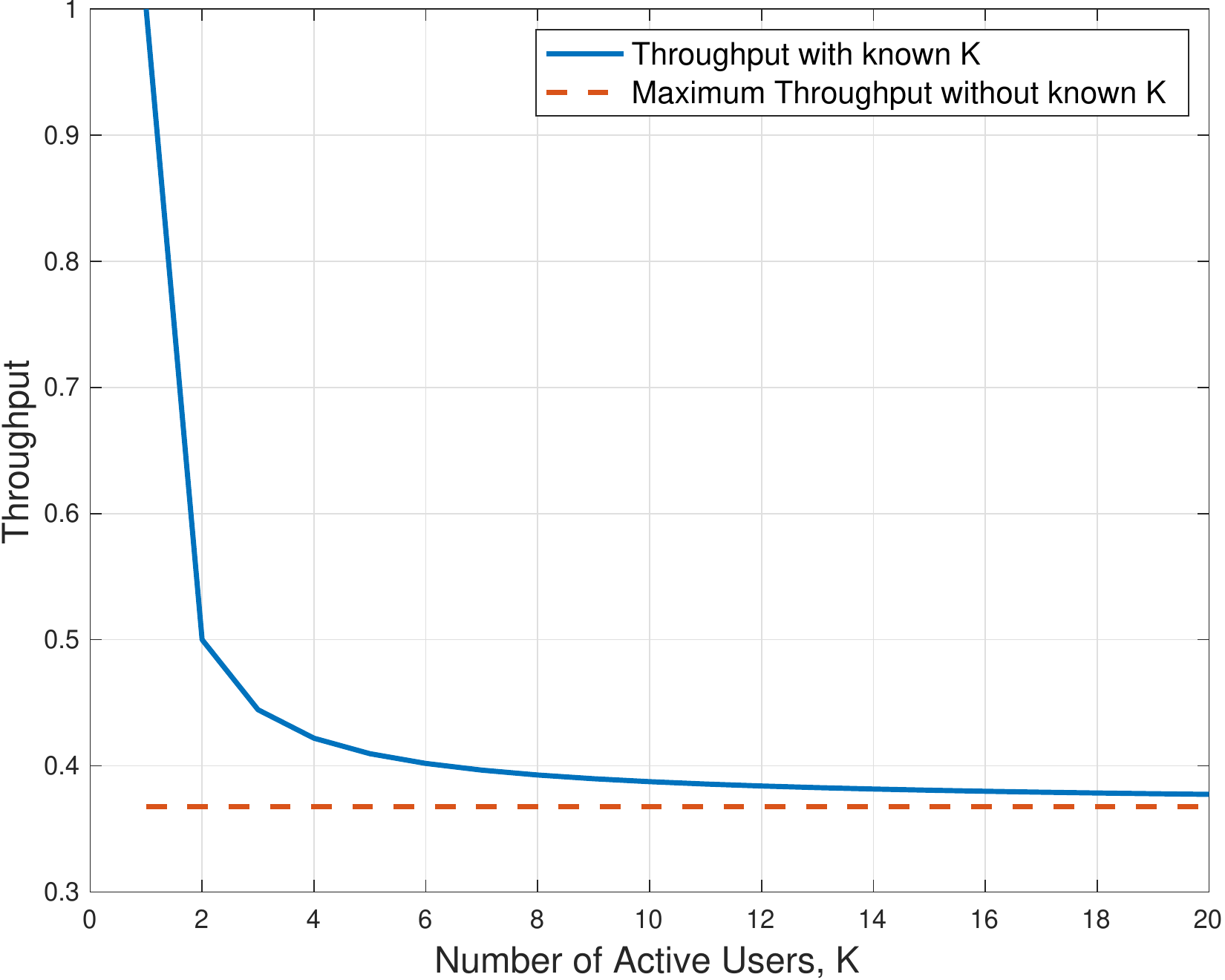}
\end{center}
\caption{Throughout with known number of active users, $K$,
and the maximum throughput without knowing $K$, i.e., $e^{-1}$.}
        \label{Fig:gap_sa}
\end{figure}

On the other hand, suppose that
the access probability, denoted by $p$, is decided without knowing $K$.
In this case, if $K$ is a Poisson random variable
with mean $\lambda$, 
where $\lambda$ is seen as the packet arrival rate,
the throughput becomes
\be
\eta_{\rm sa}  = p \lambda e^{- p \lambda} \le e^{-1},
	\label{EQ:esa_p}
\ee
where the upper-bound can be achieved by $p = \frac{1}{\lambda}$ for
$\lambda \ge 1$.
Therefore, there is a gain\footnote{The gain can be offset by
the overhead due to exploration, i.e., the overhead due
to preamble transmissions.
However, if $T_{\rm p} \ll T_{\rm d}$, the offset might be
negligible. We will discuss more details later.} 
(i.e., the difference between \eqref{EQ:esa} and \eqref{EQ:esa_p})
obtained by the exploration that
allows active users to know how many are in contention,
i.e, the number of active users, $K$.
In addition, as shown in 
\eqref{EQ:ineq}, the gain increases if $K$ is small,
which is also illustrated in Fig.~\ref{Fig:gap_sa}.

\subsection{The Proposed System based on Preamble-based Exploration}

In this subsection, we propose a 
multichannel ALOHA system with exploration using preamble transmissions.

Suppose that there are $M$ orthogonal
radio resource blocks for multiple access channels.
We assume that each active user
can randomly choose one channel and
transmit a preamble signal to the BS
in the exploration phase (EP).
After the EP, the BS can find 
the number of active users for each channel.
Let $k_m$ denote 
the number of active users transmitting their preambles
through the $m$th channel.
Then, the BS broadcasts the numbers of active
users for all the channels, $\{k_1, \ldots, k_M\}$
so that all the active users can see the state of
contention.
For example, an active user transmits a preamble
through the 1st channel and sees that $k_1 = 1$.
In this case, clearly,
the user is only one active user using 
channel 1.

Let 
$$
\cS = \{m \,|\, k_m = 1, \ m = 1,\ldots, M\},
$$
i.e., $\cS$ is the index set of the channels with only
one active user transmitting preamble.
For convenience, let $\cS^c$ denote
the complement of $\cS$.
In the data transmission phase (DTP),
if a user transmitting a preamble through 
channel $m$ sees that $k_m = 1$ or $m \in \cS$, the user
can send a data packet through the $m$th channel.
For convenience, this user is referred to as a contention-free
user. The group of contention-free active users 
is also referred to as Group I.
On the other hand, when $m \in \cS^c$ (which implies
that $k_m \ge 2$ as the user transmits a preamble
through channel $m$ and $m \notin \cS$),
the user is referred to as a user in contention, and the
group of active users in  contention is referred to as
Group II.

\begin{figure}[thb]
\begin{center}
\includegraphics[width=\figwidth]{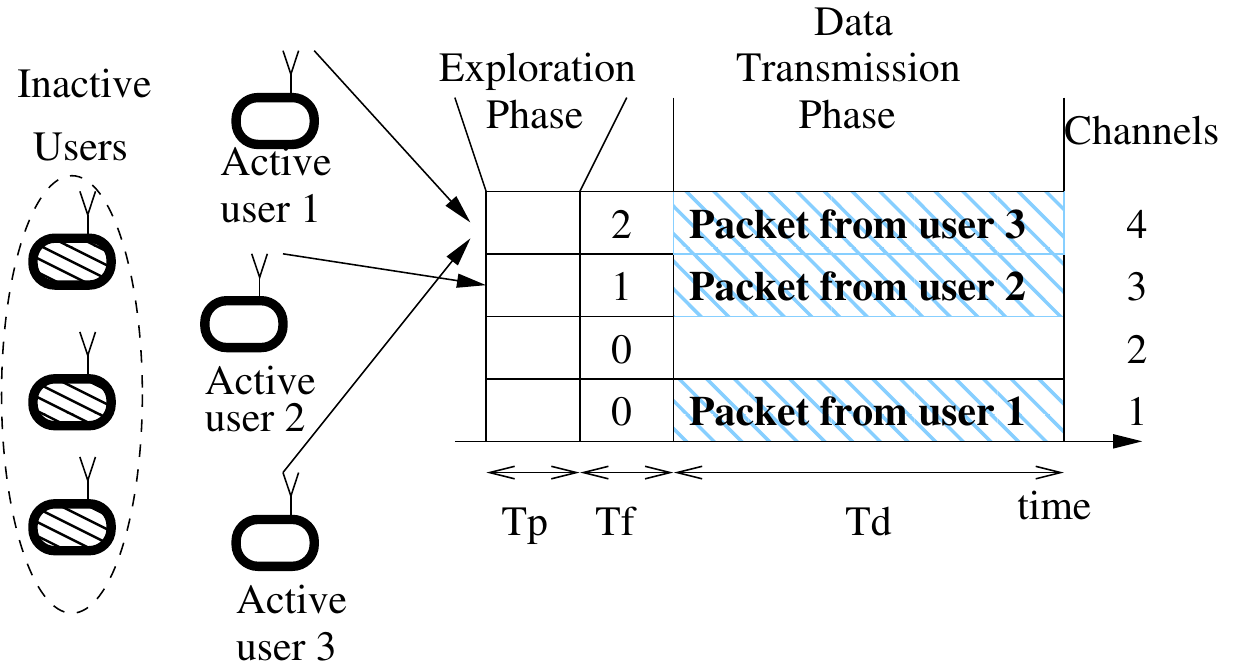}
\end{center}
\caption{An example
with $K = 3$ and $M = 4$, where active user 2 chooses
channel 3 to transmit a preamble and two other
active users (users 1 and 3) choose channel 4.}
        \label{Fig:sys}
\end{figure}

An example is shown in Fig.~\ref{Fig:sys}
with $K = 3$ and $M = 4$. Active user 2 chooses
channel 3 to transmit a preamble and two other
active users (users 1 and 3) choose channel 4.
From this, the feedback information from the BS to the
active users is $\{k_1, k_2, k_3, k_4\}
= \{0, 0, 1, 2\}$.
As a result, $\cS = \{3\}$ and $\cS^c = \{1, 2, 4\}$,
and active user 2 belongs to Group I and 
active users 1 and 3 belong to Group II.

If the user in contention transmits a packet through channel
$m$, it will be collided with others. 
To avoid packet collision, 
the user can choose a different channel, say channel $l
\in \cS^c$, and
transmits a packet. However, since another user in contention
can choose channel $l$, there can be packet collision.

Although it is not possible to prevent collisions for users
in contention, in order to mitigate packet collisions, we can assume that
each user in contention can randomly choose a channel in $\cS^c$
with an access probability, $p_{\rm dtp}$, in the DTP
to transmit a packet, while
a user in contention does not transmit a packet with
probability $1 - p_{\rm dtp}$.

It is noteworthy that 
since the feedback after EP in the proposed approach
is not used for access control,
it is different from 
existing collision-resolution algorithms
\cite{BertsekasBook}.
In the proposed multichannel ALOHA with EP,
thanks to multiple channels, the feedback after EP
is used to keep a selected
channel (in Group I) or re-choose a channel 
(in Group II) by each active user.
According to Fig.~\ref{Fig:sys},
a frame consists of EP, feedback, and DTP. After
each frame, another feedback can be transmitted
from the BS to inform success or collision of packet transmissions,
which can be used for access control as in conventional
random access schemes.

\subsection{Estimation of Number of Transmitted Preambles}
\label{SS:DNTP}

In multichannel ALOHA with EP, we have assumed 
that the BS can estimate $\{k_m\}$. In this subsection, we  
provide examples to demonstrate the estimation of $\{k_m\}$.

There can be two different cases when 
an active user transmits a preamble. 
In the first case, a common
set or pool of preambles,
denoted by $\cC =\{\bc_1, \ldots, \bc_{L}\}$, can be used.
Here, $\bc_l$ represents the $l$th preamble sequence
(of length $T_{\rm p}$) and $L$ denotes the number of the preambles
in $\cC$. Any active user
is to randomly choose one in $\cC$.
At the BS, the signal received through the $m$th channel can be
expressed as follows:
\be
\by_m = \bC \bs_m + \bn_m,
\ee
where $\bC = [\bc_1 \ \ldots \ \bc_L]$ and $\bn_m \sim \cC \cN(\b0,
N_0 \bI)$ is the background noise vector.
Here, the $l$th element of $\bs_m$, denoted by $s_{m,l}$, is given by
$s_{m,l} = \sum_{k \in \cK_{m,l}} h_k \sqrt{P_k}$,
where $h_k$ is the channel coefficient from active user $k$
to the BS, $P_k$ is the transmit power of active user $k$,
and $\cK_{m,l}$ is the index set of the active users choosing the 
$l$th preamble in the $m$th channel. 
Suppose that active users can decide their transmit powers 
to reach a target signal-to-noise ratio (SNR), i.e., 
$\frac{|h_k|^2 P_k}{N_0} \ge \Gamma$,
where $\Gamma$ is the target SNR.
It is expected that $\bs_m$ is a $k_m$-sparse vector
(recall that $k_m$ is the number 
of the active users in the $m$th channel).
Then, when $L > T_{\rm p}$, using compressive sensing algorithms 
\cite{Eldar12},
the BS is able to estimate $\bs_m$ under certain conditions
(of $\bC$ and the maximum sparsity of
$\bs_m$) \cite{Donoho06} \cite{Candes06},
which has been discussed in the context of compressive
random access \cite{Applebaum12} 
\cite{Wunder14} \cite{Schepker15} \cite{Choi17IoT}.
Once $\bs_m$ is estimated, 
the determination of the number of active users 
or the estimation of $k_m$ becomes 
straightforward (because $k_m$ can be found from the sparsity
of $\bs_m$),
while preamble collision\footnote{It can happen
as a preamble can be chosen by multiple active users.}
\cite{Choi18c} \cite{Seo19}
can result in errors in estimating $k_m$.
From \cite{Mitz05},
for the $m$th channel, 
the conditional probability of no preamble collision
can be found as
\begin{align}
\uP_m (k_m)
& = \prod_{k=1}^{k_m-1} \left(1 - \frac{k}{L} \right) 
\approx e^{- \frac{k_m (k_m-1)}{2 L}},
\end{align}
where the approximation is actually a lower-bound
(thus, $1 - e^{- \frac{k_m (k_m-1)}{2 L}}$ is an upper-bound
on the conditional probability of preamble collision).
If $K$ active users are uniformly distributed
over $M$ channels and $K$ is assumed to follow a Poisson 
distribution with mean $\lambda$,
it can be shown that
\begin{align}
\uP_m 
& = \sum_{k_m = 0}^\infty e^{- \frac{k_m(k_m-1)}{2 L}}
p_{\frac{\lambda}{M}} (k_m) \cr
& \approx
\sum_{k_m = 0}^\infty 
\left(1- \frac{k_m(k_m-1)}{2 L} \right) p_{\frac{\lambda}{M}} (k_m) 
%& = 1 - \frac{1}{2L}
%\sum_{k_m = 2}^\infty  k_m (k_m-1) p_{\frac{\lambda}{M}} (k_m) \cr
= 1 - \frac{\lambda^2}{2 L M^2}, \quad
\end{align}
where $\frac{\lambda}{M} < 1$.
Thus, $\frac{\lambda^2}{2 L M^2}$ 
becomes the probability of preamble collision.
It can be shown that
\be
1 - \uP_m 
\le \delta \Rightarrow  \frac{\lambda^2}{2 L M^2} \le \delta,
\ee
where $\delta \ll 1$ is a threshold probability of preamble collision.
To keep $\delta$, we need
\be
L \ge \frac{\lambda^2}{2 \delta M^2}.
	\label{EQ:Lge}
\ee

In the second case, it is assumed that all users
have unique preamble sequences. In this case, there is no preamble collision.
However, there are
a large number of columns of the matrix of preambles
(which is the same as the number of all users),
which makes the sparse signal estimation (and
the determination of the number of active users) difficult.
To avoid it, as suggested in \cite{Choi_18Feb},
sparse preamble sequences can be used.

In summary, by exploiting the notion of compressive sensing,
it is possible to determine $k_m$ at the BS. 
Note that, in the first case,
it is not necessary that the preamble sequences 
are orthogonal. For example, we can use
Zadoff-Chu or Alltop sequences \cite{Foucart13} for
preambles with reasonably low cross-correlation. 
In this 
case, the number of
preambles becomes $L = T_{\rm p}^2$ (for Alltop sequences)
when $T_{\rm p} \ge 5$ is a prime.
That is, a large number of preambles
to keep the probability of preamble collision 
low can be obtained with a reasonable length of preamble, $T_{\rm p}$.

\section{Maximum Throughput Comparison}	\label{S:MT}

In this section, we focus on the maximum throughput
for two schemes, namely conventional multichannel ALOHA
and multichannel ALOHA with EP.
As mentioned earlier, it is assumed that
the length of data packet, $T_{\rm d}$, is the same
for all active users. 
We show that the ratio of the maximum
throughput of 
multichannel ALOHA with EP to that of
conventional multichannel ALOHA becomes $2 - e^{-1} \approx 1.632$,
which can be seen as the gain of exploration.

\subsection{Throughput of Conventional Multichannel ALOHA}

In conventional multichannel ALOHA,
we can find the 
average number of packets without collisions
as follows:
\be
N_{\rm ma} (K,M) = K \left(1 - \frac{1}{M} \right)^{K-1}.
\ee
For a large $K$, it can be shown that
\begin{align}
N_{\rm ma} (K,M)
& = K \left(1 - \frac{1}{M} \right)^{K-1} \approx K e^{- \frac{K}{M}} \cr
& \le \bar N_{\rm ma} (M) = M e^{-1},
	\label{EQ:ma}
\end{align}
where $\bar N_{\rm ma}(M)$ is the maximum throughput
of multichannel ALOHA,
which can be achieved if $K = M$.
In addition, as in \cite{Shen03},
the arrival rate has to be lower than $M e^{-1}$ for 
system stability.

\subsection{Throughput of Multichannel ALOHA with EP}

One of the main results 
of the paper can be stated as follows.

\begin{mytheorem}	\label{T:1}
The ratio of the maximum throughput of  multichannel ALOHA with EP 
to that of conventional multichannel ALOHA is
\be
\eta = 2 - e^{-1}.
	\label{EQ:T1}
\ee
\end{mytheorem}
In the rest of this subsection, we prove Theorem~\ref{T:1}.

Prior to the proof, we can briefly compare 
the normalized maximum throughput 
(per channel) with those of well-known schemes as follows:
\begin{align*}
\max \frac{N_{\rm ma}}{M} & = e^{-1} \approx 0.3679 \cr
\max \frac{N_{\rm ep}}{M} & = e^{-1} (2 - e^{-1}) \approx 0.6004 \cr
\max \frac{N_{\rm co}}{M} & = 1,
\end{align*}
where $N_{\rm co}$ represents the throughput of coded random access.
As in \cite{Paolini15}, the
normalized throughput of coded random access can approach
1 under ideal conditions
(i.e., all the transmitted packets can be recovered
regardless of collisions).
Furthermore, for NOMA-based random
access, the maximum normalized throughput
can be higher than 1 thanks to virtual multiple channels
in the power-domain. However, both 
coded and NOMA-based random access schemes need to have SIC
at the BS. On the other hand, the proposed scheme
(i.e., multichannel ALOHA with EP) does not require SIC
like conventional multichannel ALOHA.

Let $S = |\cS|$,
where $S$ represents the number of active users
that can transmit packets without collisions.
Clearly, $S \le \min\{K, M\}$.
In addition, let $W = K - S$, where
$W$ becomes the number of active users in contention.
For convenience, let
\be
L = M - S.
\ee
Clearly, $L$ is the number of the channels that are available
for contention-based transmissions for $W$ active users in contention
or Group II.
In addition, suppose that among $W$,
$U$ active users in contention are to transmit 
their packets through $L$ channels.
Clearly, for given $W$, $U$ has the following  
distribution:
\be
\uP(U = u\,|\,W) = \binom{W}{u} p_{\rm dtp}^u (1 - p_{\rm dtp})^{W-u}.
	\label{EQ:Pu}
\ee
Throughout the paper, we assume that
\be
p_{\rm dtp} = \min \left\{
1, \frac{L}{W} \right\},
	\label{EQ:p_dtp}
\ee
which maximizes\footnote{For Group II, we can consider
a multichannel ALOHA system with $L$ channels and $W$ users. 
In this case, $p_{\rm dtp}$ is seen
as the access probability that can maximize
the throughput if it is given in \eqref{EQ:p_dtp}
\cite{Shen03}, \cite{Chang15}.}
the average number of 
packets without collisions in Group II.

For a given $U$,
the conditional average number of the packets that can be successfully
transmitted from $U$ users without collisions in contention during DTP
becomes $U \left(1 - \frac{1}{L} \right)^{U-1}$.
As a result, the 
conditional average number of packets without collisions
(for given $U$, $L$, and $S$)
is given by
\be
N_{\rm ep} (U, L, S) =  S + U \left(1 - \frac{1}{L} \right)^{U-1},
	\label{EQ:VULS}
\ee
where the first term on the right-hand side (RHS) in \eqref{EQ:VULS}
is the number of packets without collisions from Group I
and the second term is that from Group II.
For convenience, let 
\be
N_{\rm ep} (K,M) = \uE[N_{\rm ep} (U, L, S) \,|\, K],
\ee
which is the average number of 
packets without collisions (for given $K$ and $M$) in multichannel ALOHA
with EP.

\begin{mylemma}
Suppose that $M$ and $K$ are sufficiently large
so that $L$ and $W$ are also large.
The upper bound on 
$N_{\rm ep} (K,M)$ is given by
\be
N_{\rm ep} (K,M)  \le M e^{-1} + \bar S (K) (1 - e^{-1}).
	\label{EQ:L1}
\ee
\end{mylemma}
\begin{IEEEproof}
From Eq.~\eqref{EQ:VULS}, we have
\begin{align}
N_{\rm ep} (K,M) & = \uE[ N_{\rm ep} (U, L, S) \,|\, K] \cr
& =  \uE[ S\,|\,K] + 
\uE\left[U \left(1 - \frac{1}{L} \right)^{U-1}\,|\, K
\right].
	\label{EQ:Vep1}
\end{align}
In \eqref{EQ:Vep1}, it can be shown that
\begin{align}
\bar S(K) = \uE[S \,|\, K] 
= K \left(1 - \frac{1}{M} \right)^{K-1}.
	\label{EQ:bSK}
\end{align}
Since $U$ depends on $W$ and $W$ depends on $K$,
we have
$$
\uE \left[  U \left(1 - \frac{1}{L} \right)^{U-1}
 \,\bigl|\,  K \right]
=
\uE \left[ \left[ U \left(1 - \frac{1}{L} \right)^{U-1}
 \,\bigl|\, W \right]\,\bigl|\, K \right]
$$
From \eqref{EQ:Pu},
after some manipulations,
it can be shown that
\begin{align}
\uE\left[ U \left(1 - \frac{1}{L} \right)^{U-1}
 \,|\, W \right] 
%& = \sum_{u=0}^W \uP(U = u \,|\, W) 
%u \left(1 - \frac{1}{L} \right)^{u-1} \cr
& = p_{\rm dtp} W \left(
1 - \frac{p_{\rm dtp}}{L} \right)^{W-1}.
\end{align}
If $\frac{p_{\rm dtp}}{L} \ll 1$,
it follows that
$p_{\rm dtp} W \left(
1 - \frac{p_{\rm dtp}}{L} \right)^{W-1}
\approx 
p_{\rm dtp} W e^{-\frac{p_{\rm dtp} W}{L}}
\le L e^{-1}$.
The upper bound can be achieved if 
\be
p_{\rm dtp} = \frac{L}{W} \le 1.
	\label{EQ:op}
\ee
As a result, since $L = M - S$, it can be shown that
\begin{align}
\uE \left[  U \left(1 - \frac{1}{L} \right)^{U-1}
 \,\bigl|\,  K \right]
& \le \uE[L e^{-1}\,|\, K] \cr
& = (M - \uE[S\,|\, K])e^{-1}.
	\label{EQ:EU}
\end{align}
Substituting \eqref{EQ:bSK} and \eqref{EQ:EU}
into \eqref{EQ:Vep1}, we have
\begin{align}
N_{\rm ep} (K,M) & \le \bar S(K) +(M - \bar S(K) ) e^{-1} \cr
& = M e^{-1} + \bar S(K) (1 - e^{-1}), 
\end{align}
which completes the proof.
\end{IEEEproof}

From \eqref{EQ:L1},
the maximum throughput of multichannel ALOHA with EP 
is given by 
\begin{align}
\bar N_{\rm ep} (M) & = \max_K N_{\rm ep} (M,K) \cr
& = M e^{-1} + (1- e^{-1}) \max_K \bar S(K).
\end{align}
For a sufficiently large $K$, $\bar S(K) \approx K e^{-\frac{K}{M}}$. 
Thus, if we consider the throughput gain
using the ratio of the maximum throughput of multichannel ALOHA
with EP to that of conventional ALOHA, it can be shown that
\begin{align}
\eta & = \frac{\bar N_{\rm ep} (M)}{\bar N_{\rm ma} (M)} 
= \frac{M e^{-1} + (1- e^{-1}) M e^{-1}}{M e^{-1}} \cr
& = 2 - e^{-1} \approx 1.632,
	\label{EQ:eta}
\end{align}
which finally proves Theorem~\ref{T:1}.
From this, it is clear that
the EP can improve the performance
of multichannel ALOHA (in terms of the throughput) by a factor of 1.632.

\section{Steady-State Analysis with Fast Retrial}	\label{S:SS}

In this section, a steady-state analysis
is carried out when fast retrial in 
\cite{YJChoi06} is used. The throughput and probability of collision
are found when the total arrival rate
is modeled as a Poisson random variable.

\subsection{Steady-State Analysis of Conventional Multichannel ALOHA}
\label{SS:S1}

Suppose that the number of new arrivals follows a Poisson 
distribution with mean $\lambda_0$. 
As in \cite{YJChoi06},
we assume that the total of the new and back-logged arrivals,
which is $K$, follows a Poisson distribution 
with mean $\lambda$ that is given by
\be
\lambda = \lambda_0 + q_{\rm ma} \lambda,
	\label{EQ:lamlam}
\ee
where $q_{\rm ma}$ is the probability of collision of
conventional multichannel ALOHA.

For a given $\lambda$,
under the assumption that $K$ is a Poisson
random variable with mean $\lambda$,
the average number of packets without collisions becomes
\begin{align}
N_{\rm ma} (M) & = 
\uE\left[ K \left( 1 - \frac{1}{M} \right)^{K-1} \right] \cr
& = \sum_{k=0}^\infty k \left( 1 - \frac{1}{M} \right)^{k-1} 
p_\lambda (k) 
= \lambda e^{-\frac{\lambda}{M}},
\end{align}
where $p_\lambda (k) = \frac{e^{-\lambda} \lambda^k }{k!}$
is the probability mass function of a Poisson random
variable with mean $\lambda$. 
As in \cite{YJChoi06}, $q_{\rm ma}$ can be given by
the following ratio:
\begin{align}
q_{\rm ma} 
& = \frac{\uE[\mbox{Number of collided packets}]}
{\uE[\mbox{Number of transmitted packets}]} \cr
& = 1 - \frac{N_{\rm ma} (M) }{\uE[K] } 
= 1  - e^{- \frac{\lambda}{M}}.
	\label{EQ:qma}
\end{align}
Then, from \eqref{EQ:lamlam} and \eqref{EQ:qma},
we can see that
$\lambda$ is the solution of the following equation:
\be
\lambda_0 = \lambda e^{- \frac{\lambda}{M}}, \ \lambda \in [0, M),
	\label{EQ:cll}
\ee
Furthermore, it can be shown that
$N_{\rm ma} (M) = \lambda_0$,
which means that in the steady state, the input
arrival rate, i.e., 
$\lambda_0$, is identical to the output rate
that is
the average number of packets without collisions.

\subsection{Steady-State Analysis of Multichannel ALOHA with EP}
\label{SS:2}

For given $K$ and $M$,
the number of packets without collisions 
can be re-written as
\begin{align}
N_{\rm ep} (K,M) = \uE \left[S + \sum_{w = 1}^W Z_w \,|\, K \right],
	\label{EQ:NKM0}
\end{align}
where $Z_w \in \{0,1\}$ is 1 if the $w$th 
active user in Group II transmits a packet, and otherwise
it becomes 0.
For steady-state analysis, as in Subsection~\ref{SS:S1}, we need to find 
$\uE[N_{\rm ep} (K,M)]$, where the expectation 
is carried out over $K$.
Unfortunately, it is not easy to find a closed-form expression for
$\uE[N_{\rm ep} (K,M)]$.
Thus, we resort to a bound as follows.

\begin{mytheorem}
Suppose that $K$ follows a Poisson distribution
with mean $\lambda$. Then, a lower-bound on the average
number of packets without collisions
is given by
\begin{align}
N_{\rm ep} (M) 
& = \uE[N_{\rm ep} (K,M)] \ge (1-e^{-1}) \lambda e^{-\frac{\lambda}{M}} \cr
&\ + e^{-1} \left( 
M (1- F_\lambda(M)) +\lambda F_\lambda(M-1) 
\right),
	\label{EQ:lb}
\end{align}
where $F_\lambda(M) = \sum_{m=0}^M p_\lambda(m)$ 
is the cumulative
distribution function 
of a Poisson random variable
with mean $\lambda$.
\end{mytheorem}
\begin{IEEEproof}
In \eqref{EQ:NKM0}, it can be shown that
\begin{align}
\uE[Z_w\,|\, W, L] = p_{\rm dtp} 
\left(1 - \frac{p_{\rm dtp}}{L} \right)^{W-1}. 
\end{align}
Suppose that $K \le M$. In this case, 
we have $\frac{L}{W} = \frac{M-S}{K-S} \ge 1$, $p_{\rm dtp}$ has to be
1 according to \eqref{EQ:p_dtp}. 
Thus,
it can be shown that
\be
\uE\left[
\sum_{w = 1}^W Z_w \,|\, W,L\right] 
= W \left(1 - \frac{1}{L} \right)^{W-1}. 
	\label{EQ:WZ}
\ee
Substituting \eqref{EQ:WZ}
into \eqref{EQ:NKM0}, we have 
\begin{eqnarray}
N_{\rm ep} (K,M)
%& = & \uE \left[S + (K-S) \left(1 - \frac{1}{M-S} \right)^{K-1 - S}
%\, \bigl|\, K \right] \cr
%& \ge & \uE \left[S + (K-S) \left(1 - \frac{1}{M-S} \right)^{M-1- S} 
%\, \bigl|\, K \right] \cr
\ge K e^{-1} + \uE[S\,|\, K] (1- e^{-1}).
	\label{EQ:KM1}
\end{eqnarray}

We now consider the case that $K > M$.
It can be shown that
\begin{align}
N_{\rm ep} (K,M) & = 
\uE \left[S + W p_{\rm dtp}
\left(1 - \frac{p_{\rm dtp}}{L} \right)^{W-1} \,\bigl|\, K \right] \cr
%& = \uE \left[
%S+ L \left(1 - \frac{1}{W} \right)^{W-1}\,\bigl|\, K \right] \cr
& \ge \uE[S+ L e^{-1} \,|\, K] \cr
& = M e^{-1} + \uE[S\,|\, K] (1-e^{-1}).
	\label{EQ:KM2}
\end{align}

From \eqref{EQ:KM1} and \eqref{EQ:KM2},
we have
\begin{align}
N_{\rm ep} (K,M) \ge \min\{K,M\} e^{-1} + \uE[S\,|\,K] (1 - e^{-1}).
\end{align}
Since $K$ is a Poisson random variable with mean $\lambda$,
using \eqref{EQ:bSK},
it can be shown that
\begin{align}
\uE[N_{\rm ep} (K,M)] & \ge 
e^{-1} \sum_{m=0}^\infty \min\{m,M\} p_\lambda (m) \cr
& \ + (1 - e^{-1}) 
\sum_{m=0}^\infty
m\left(1 - \frac{1}{M} \right)^{m-1} p_\lambda(m) \cr
& =
e^{-1} \sum_{m=0}^\infty \min\{m,M\} p_\lambda (m) \cr
& \ + (1 - e^{-1}) \lambda e^{-\frac{\lambda}{M}}.
	\label{EQ:NKM}
\end{align}
After some manipulations, we have
\begin{align}
& \sum_{m=0}^\infty \min\{m,M\} p_\lambda (m) \cr
& = \sum_{m=0}^\infty m p_\lambda (m) 
- \sum_{m=M+1}^\infty (m -M) p_\lambda (m)  \cr
& = \lambda - (\lambda - M) (1 - F_\lambda(M-1)) - 
M p_\lambda (M).
	\label{EQ:minp}
\end{align}
Substituting \eqref{EQ:minp} into \eqref{EQ:NKM},
we can obtain the RHS term in \eqref{EQ:lb},
which completes the proof.
\end{IEEEproof}

Suppose that $\lambda$ is sufficiently smaller
than $M$ so that $|1 - F_\lambda (M)|,
|1 - F_\lambda (M-1)| \le \epsilon$,
where $0 < \epsilon \ll 1$.
Then, from \eqref{EQ:lb}, we can show that
\be
N_{\rm ep} (M) \ge 
(1 - e^{-1}) \lambda e^{-\frac{\lambda}{M}}
+ e^{-1} \lambda + O(\epsilon).
\ee
Since $\lambda \ge \lambda e^{-\frac{\lambda}{M}}$,
it can be shown that
$(1 - e^{-1}) \lambda e^{-\frac{\lambda}{M}}
+ e^{-1} \lambda \ge \lambda e^{-\frac{\lambda}{M}}$.
Thus, ignoring the term of $O(\epsilon)$, it can be shown
that 
\be
N_{\rm ep} (M) - N_{\rm ma} (M)
\ge e^{-1} \lambda \left(1 - e^{-\frac{\lambda}{M}}\right),
	\label{EQ:NNineq}
\ee
which shows that 
the average number of packets without collisions
of multichannel ALOHA with EP is larger than
that of conventional multichannel ALOHA
and the performance gap increases as $\lambda$ increases.
Furthermore, since \eqref{EQ:NNineq} is re-written as
\be
N_{\rm ep} (M) - N_{\rm ma} (M)
\ge e^{-1} \alpha (1 - e^{-\alpha}) M,
	\label{EQ:Mgap}
\ee
where $\alpha = \frac{\lambda}{M}$,
with a fixed $\alpha$, we can see that the gap grows linearly
with $M$.

In general, we expect that $\lambda < M$
to avoid a large number of back-logged packets.
In this case, an approximation of $N_{\rm ep} (M)$
can be found as follows.

\begin{mylemma}
Suppose that $M$ is sufficiently large. If $\lambda < M$,
$N_{\rm ep} (M)$ can be approximated by
\be
N_{\rm ep} (M)
\approx \tilde N_{\rm ep}(M) = \lambda - \frac{A_1 - 2 A_2 + A_3}{M},
	\label{EQ:L2}
\ee
where 
$A_1 =  \lambda (1+\lambda)$,
$A_2 =  \lambda e^{-\frac{\lambda}{M}} 
\left(1+\lambda \left(1 - \frac{1}{M} \right) \right)$, and 
$A_3 =  \lambda e^{-\lambda + \lambda \left(1-\frac{1}{M}\right)^2} 
\left(1+\lambda \left(1 - \frac{1}{M} \right)^2 \right)$.
\end{mylemma}
\begin{IEEEproof}
Since $\lambda < M$, 
$N_{\rm ep} (K,M)$ can be approximated by \eqref{EQ:KM1}
(i.e., for the case of $K \le M$).
In addition, in \eqref{EQ:KM1},
we can show that
\begin{align}
& S + (K-S) \left(1 - \frac{1}{M-S}\right)^{K-1-S} \cr
& \approx S +  (K-S) \left(1 - \frac{1}{M}\right)^{K-1-S} 
\ (\mbox{for $M \gg S$}) \cr
& \approx S +  (K-S) \left(1 - \frac{K-S}{M}\right) 
\ (\mbox{for $M \gg 1$}) \cr
& = K - \frac{(K-S)^2}{M}.
\end{align}
Thus, an approximation of $N_{\rm ep} (M)$ is given by
\begin{align}
N_{\rm ep} (M) & \approx \uE[K] - \frac{\uE[(K-S)^2]}{M} \cr
& = \lambda -
\frac{1}{M} \uE[\uE[K^2 - 2 S K + S^2 \,|\, K]].
	\label{EQ:as1}
\end{align}
After some manipulations, we can show that
\begin{align}
\uE[K^2] & = A_1 \cr
\uE[\uE[S K\,|\, K]] & = \uE
\left[ K^2 \left(1 - \frac{1}{M} \right)^{K-1} \right] = A_2 \cr
\uE[\uE[S^2\,|\, K]] & = \uE
\left[ K^2 \left(1 - \frac{1}{M} \right)^{2(K-1)} \right] = A_3.
	\label{EQ:as2}
\end{align}
Substituting \eqref{EQ:as2} into \eqref{EQ:as1},
we can obtain \eqref{EQ:L2},
which completes the proof.
\end{IEEEproof}

\begin{mytheorem}
Suppose that $\lambda$ and $M$ increase with a fixed ratio
$\alpha < 1$.
Then, the asymptotic normalized throughput can be given by
\be
\lim_{M \to \infty} \frac{\tilde N_{\rm ep} (M)}{M}
= \psi(\alpha) =  \alpha - \alpha^2 (1 - e^{- \alpha})^2.
	\label{EQ:psi}
\ee
The function $\psi(\alpha)$ is a concave function of 
$\alpha \in (0,1)$ and has a unique maximum.
\end{mytheorem}
\begin{IEEEproof}
Since 
$\lim_{M \to \infty} \frac{A_1}{M^2} = \alpha^2$,
$\lim_{M \to \infty} \frac{A_2}{M^2} = \alpha^2 e^{-\alpha}$, and
$\lim_{M \to \infty} \frac{A_3}{M^2} = \alpha^2 e^{- 2 \alpha}$,
it can be easily shown that
\begin{align}
\lim_{M \to \infty} \frac{\tilde N_{\rm ep} (M)}{M}
= \alpha  - \alpha^2 (1 - 2 e^{-\alpha} + e^{-2 \alpha}),
\end{align}
which becomes $\psi(\alpha)$ in 
\eqref{EQ:psi}. 

The 2nd derivative of $\psi(\alpha)$ is given by
\begin{align*}
\psi^{\prime \prime}(\alpha)
& = 
- 2 \left( (1-(1-\alpha)e^{-\alpha}  )^2
+ \alpha (1- e^{-\alpha}) (2 - \alpha) e^{-\alpha}
\right) \cr
& \le 0, \ \alpha \in (0,1),
\end{align*}
which shows that $\psi(\alpha)$ is a concave 
function of $\alpha \in (0,1)$.
This completes the proof.
\end{IEEEproof}

The maximum of $\psi(\alpha)$ lies between 0 and 1
as $\psi^\prime(0) = 1 > 0$ and $\psi^\prime(1) = 2 e^{-1} - 1 < 0$, 
which is given by
\be
\max_{0 \le \alpha \le 1} \psi (\alpha) = 0.6149.
\ee
Note that the normalized maximum throughput 
of the conventional multichannel ALOHA is 
\be
\max_\lambda \frac{\lambda e^{-\frac{\lambda}{M}}}{M} 
=\max_\alpha \alpha e^{-\alpha} = e^{-1} = 
0.3679.
\ee
Clearly, when fast retrial is employed,
multichannel ALOHA
with EP can have a higher throughput than 
the conventional multichannel ALOHA and the performance
gain (in terms of the throughput ratio) is 
$\frac{\max_{0 \le \alpha \le 1} \psi (\alpha)}{e^{-1}} = 
\frac{0.6149}{0.3679} \approx 1.6715$,
which is similar to 
the ratio of the maximum throughput 
of multichannel ALOHA with EP 
to that of conventional ALOHA
in \eqref{EQ:eta}.

On the other hand,
if the system is lightly loaded,
i.e., $\lambda_0 \ll M$, the throughput
difference between the two systems is not significant.
To see this, let $\alpha_0 = \frac{\lambda_0}{M}$,
which is the normalized throughput 
or throughput per channel.
Then, from \eqref{EQ:cll}, in conventional multichannel ALOHA,
we have
\be
\alpha_0 = \alpha e^{-\alpha} \approx \alpha 
\ (\mbox{for $\alpha_0 \ll 1$}).
\ee
Using $\tilde N_{\rm ep} (M)$ as an approximation of 
$N_{\rm ep} (M)$, for given $\lambda_0$, 
from \eqref{EQ:lamlam} and \eqref{EQ:psi} (for a large $M$),
it can be shown that
\begin{align}
\alpha_0 
& =  \frac{\tilde N_{\rm ep} (M)}{M} = 
\alpha - \alpha^2 (1 - e^{-\alpha})^2 \cr
& \approx \alpha - \alpha^4 \approx \alpha 
\ (\mbox{for $\alpha_0 \ll 1$}).
	\label{EQ:l0}
\end{align}
However, the exploration gain can lead to an improvement of
access delay performance. To see this,
let $q_{\rm ep}$ be the probability of collision of
multichannel ALOHA with EP. 
From $q_{\rm ep}$, the outage probability
of access delay with $D$ re-transmissions can be given by
\begin{align}
P_{\rm ep} (D) & = \Pr(\mbox{access delay} > D) \cr
& = 1 - \sum_{d=0}^{D-1} (1 - q_{\rm ep}) q^d_{\rm ep} 
= q_{\rm ep}^D,
	\label{EQ:Dout}
\end{align}
while the average access delay is $\frac{1}{1 - q_{\rm ep}}$.
As in \eqref{EQ:lamlam},
it can be shown that
\begin{align}
q_{\rm ep} = 1 - \frac{N_{\rm ep} (M)}{\lambda}.
	\label{EQ:qN}
\end{align}
Then, for a large $M$, using $\tilde N_{\rm ep}(M) \approx N_{\rm ep} (M)$,
it can be shown that
\be
q_{\rm ep} = 1 - \frac{\psi(\alpha)}{\alpha}
= \alpha (1 - e^{-\alpha})^2,
\ee
while
\be
q_{\rm ma} = 1 - e^{-\alpha}.
\ee
Then, when $\alpha \approx \alpha_0$ is
sufficiently small, it can be shown that
$q_{\rm ep} \approx \alpha (1 - (1- \alpha))^2 
\approx \alpha_0^3$ and 
$q_{\rm ma} \approx 1 - (1- \alpha) \approx \alpha_0$.
It demonstrates that the delay outage probability
of multichannel ALOHA can be 
significantly lowered by exploration
when the system is lightly loaded.
For example, if $\alpha_0 = 0.1$ and the access 
delay threshold is $D = 2$, according to \eqref{EQ:Dout},
the delay outage probability of 
multichannel ALOHA with EP becomes $0.1^6 = 10^{-6}$,
while that of 
conventional multichannel ALOHA is $0.1^2 = 10^{-2}$.

\begin{figure}[thb]
\begin{center}
\includegraphics[width=\figwidth]{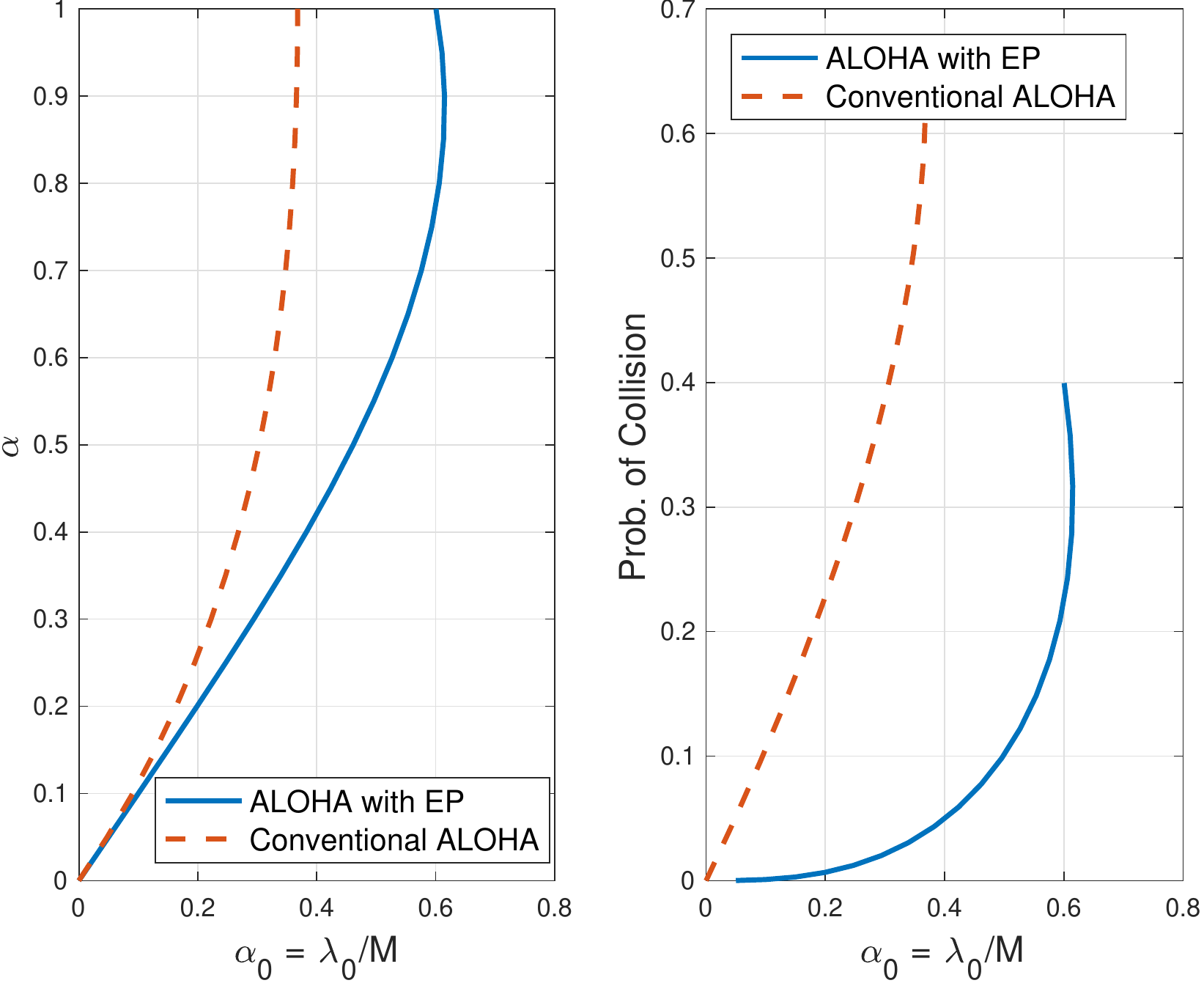} \\
\hskip 0.8cm (a) \hskip 3.5cm (b) \\
\end{center}
\caption{Comparisons between 
multichannel ALOHA with EP and conventional multichannel ALOHA:
(a) $\alpha = \frac{\lambda}{M}$ versus
$\alpha_0 = \frac{\lambda_0}{M}$;
(b) probabilities of collision as functions of
$\alpha_0 = \frac{\lambda_0}{M}$.}
        \label{Fig:aa}
\end{figure}

In Fig.~\ref{Fig:aa} (a), we show the
relationship between $\alpha_0 = \frac{\lambda_0}{M}$ and $\alpha$
for multichannel ALOHA with EP (from \eqref{EQ:l0})
and that of conventional multichannel ALOHA.
For a given normalized new arrival rate (i.e.,
$\alpha =\frac{\lambda}{M}$), 
multichannel ALOHA with EP has a lower normalized total arrival
rate (i.e., $\alpha$) than
conventional multichannel ALOHA,
which means that the probability of collision
in multichannel ALOHA with EP is lower than that in 
conventional multichannel ALOHA,
which is illustrated in Fig.~\ref{Fig:aa} (b).
When $\alpha < 0.1$ (i.e.,
a lightly loaded case), we see that $\alpha \approx \alpha_0$
in 
Fig.~\ref{Fig:aa} (a) as expected, while
$q_{\rm ep}$ is significantly lower than $q_{\rm ma}$
in Fig.~\ref{Fig:aa} (b) (which implies a 
significant decrease of the delay outage probability).
Note that the expression in \eqref{EQ:l0} from \eqref{EQ:psi}
is valid when $\alpha < 1$. Thus, as $\alpha$ approaches 1,
it cannot be used (which will also be discussed
in Section~\ref{S:Sim}).

\section{Implementation Issues}	\label{S:Imp}

\subsection{Cost for Exploration}

In multichannel ALOHA with EP, 
each active user has to transmit 
a preamble prior to packet transmission. 
Recalling that $T_{\rm f}$ is 
the length of feedback signal,
the length of slot in multichannel ALOHA with EP is 
$T_{\rm p}+ T_{\rm d} + 2 T_{\rm f}$,
while that in 
conventional ALOHA is $T_{\rm d} + T_{\rm f}$.
In  multichannel ALOHA with EP,
there are two types of feedback: one is for the numbers of transmitted
preambles in $M$ channels and the other 
is for the collisions of packets from the active 
users in Group II.
Thus, the following factor can be considered:
\be
\kappa = 
\frac{T_{\rm d} + T_{\rm f}}
{T_{\rm p}+ T_{\rm d} + 2 T_{\rm f}} = \frac{1}{1 + \epsilon_T},
\ee
where $\epsilon_T = \frac{T_{\rm p} + T_{\rm f}}{T_{\rm d} + T_{\rm f}}$.
The effective throughput of 
multichannel ALOHA with EP for
the comparison with that of conventional
multichannel ALOHA becomes $\kappa N_{\rm ep} (M)$.
As a result, if 
$$
\kappa N_{\rm ep} (M) > N_{\rm ma} (M),
$$
the exploration 
becomes beneficial to improve the performance
of multichannel ALOHA.
As mentioned earlier, 
since $T_{\rm d} \gg T_{\rm p}$, we can see that
$\kappa \approx 1$. Thus, in general, 
multichannel ALOHA with EP can have a better performance than
conventional multichannel ALOHA.

\subsection{Downlink for Feedback}

As mentioned earlier, in multichannel ALOHA with EP,
the BS needs to feed back the number of active users
in each channel, $\{k_1, \ldots, k_M\}$. Thus,
if $n_{\rm f}$ bits are allocated for each $k_m$,
there might be $n_{\rm f} M$ bits required for the feedback.
Here, $n_{\rm f} = \lceil \log_2 \max k_m \rceil$, where 
$\max k_m$ might be a constant.

In fact,
the number of feedback bits can be reduced.
For each channel, it is necessary to send one bit:
$b_m = 1$ if $k_m = 1$ and $b_m = 0$ otherwise,
where $b_m$ is one-bit feedback for channel $m$.
If an active user that transmits a preamble to channel $m$ 
receives $b_m = 1$, this user belongs to Group I (i.e., contention-free),
and $\cS  = \{m\,|\, b_m = 1\}$.
Otherwise, an active user becomes a member of Group II.
In this case, to decide $p_{\rm dtp}$,
the BS needs to send additional information, which is $W$,
while $L$ can be found at any active user from $\{b_m\}$
as $L = M - \sum_{m=1}^M b_m$.
Thus, a total number of feedback bits is $M + 
\lceil \log_2 \max W \rceil$.

\section{Simulation Results}
\label{S:Sim}

In this section, we present simulation results 
for multichannel ALOHA when $K$ follows a Poisson distribution
with mean $\lambda$.
In addition, we only consider the case that 
$\lambda \le M$. Note that if $\lambda > M$,
the system is overloaded. In  
this case, since there might be more active users than
channels, it is expected that $k_m > 1$ for most $m$.
Therefore, the exploration gain would be diminished
and multichannel ALOHA with EP becomes less useful.

\begin{figure}[thb]
\begin{center}
\includegraphics[width=\figwidth]{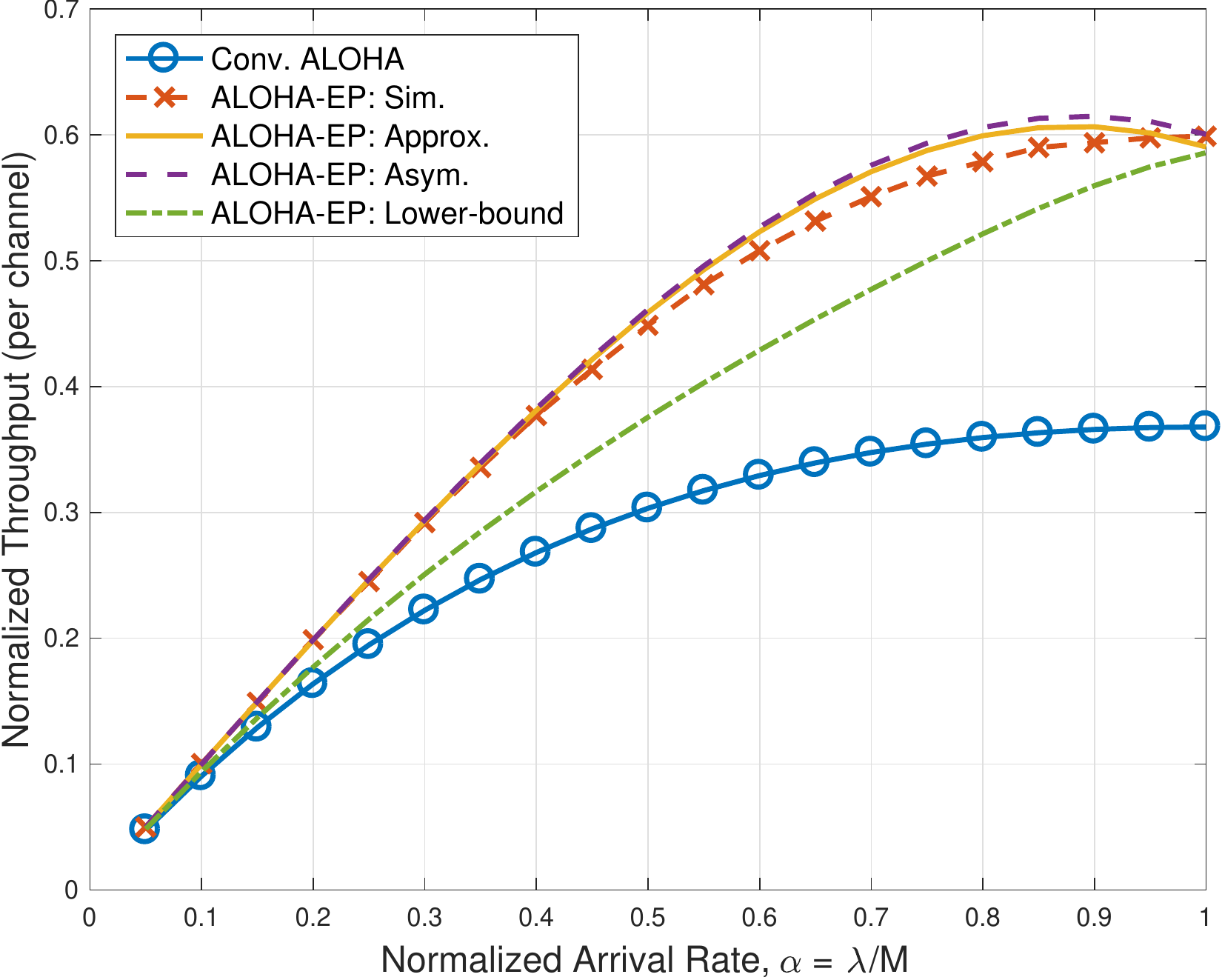}
\end{center}
\caption{Normalized throughputs, 
$\frac{\uE[N_{\rm ma}(M)]}{M}$ for
conventional multichannel ALOHA and
$\frac{\uE[N_{\rm ep}(M)]}{M}$ for
multichannel ALOHA with EP, as functions
of the normalized arrival rate, $\frac{\lambda}{M}$,
when $M = 100$.}
        \label{Fig:Tplt1}
\end{figure}

In Fig.~\ref{Fig:Tplt1},
the normalized throughputs,
$\frac{\uE[N_{\rm ma}(M)]}{M}$ for
conventional multichannel ALOHA and
$\frac{\uE[N_{\rm ep}(M)]}{M}$ for
multichannel ALOHA with EP, are shown as functions
of the normalized arrival rate, $\alpha = \frac{\lambda}{M}$,
when $M = 100$.
Clearly, we can see that 
multichannel ALOHA with EP can provide a higher throughput
than conventional multichannel ALOHA.
For multichannel ALOHA with EP, 
we also show the lower-bound and approximation
in \eqref{EQ:lb} and \eqref{EQ:L2}, respectively.
In addition, the normalized
throughput from the asymptotic throughput expression in \eqref{EQ:psi}
is presented.
When $\alpha$ is sufficiently low (e.g., 
$\alpha \le 0.5$), we can see that the
approximation in \eqref{EQ:L2} (and
the asymptotic expression in \eqref{EQ:psi}) agrees
with simulation results.

\begin{figure}[thb]
\begin{center}
\includegraphics[width=\figwidth]{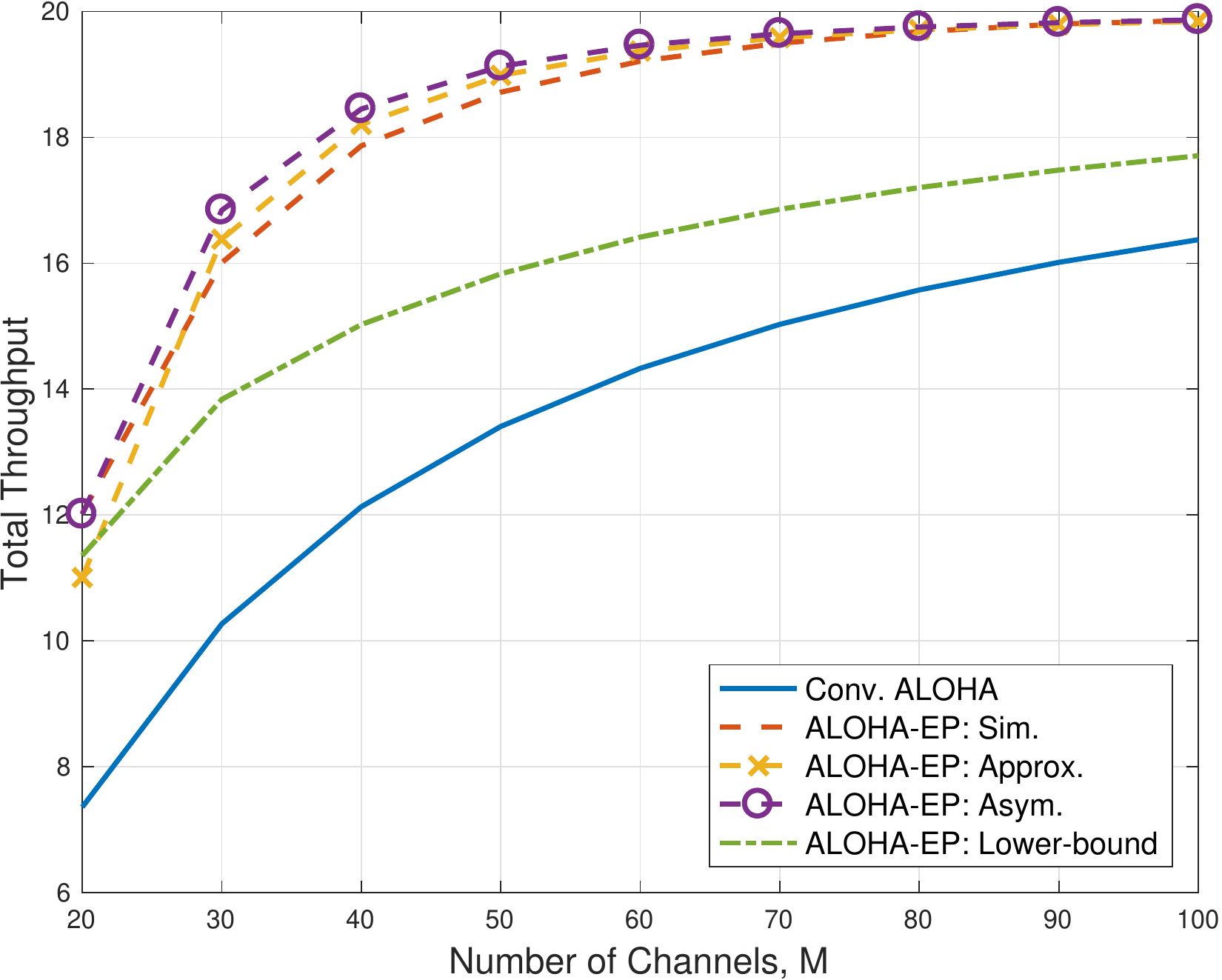}
\end{center}
\caption{Total throughputs 
of conventional multichannel ALOHA and
multichannel ALOHA with EP as functions
of $M$ when $\lambda = 20$.}
        \label{Fig:Tplt2}
\end{figure}

Fig.~\ref{Fig:Tplt2} shows the total throughputs 
of
conventional multichannel ALOHA and
multichannel ALOHA with EP as functions
of $M$ when $\lambda = 20$. As in 
Fig.~\ref{Fig:Tplt1}, we can see that the exploration
can help improve the throughput of multichannel ALOHA.

\begin{figure}[thb]
\begin{center}
\includegraphics[width=\figwidth]{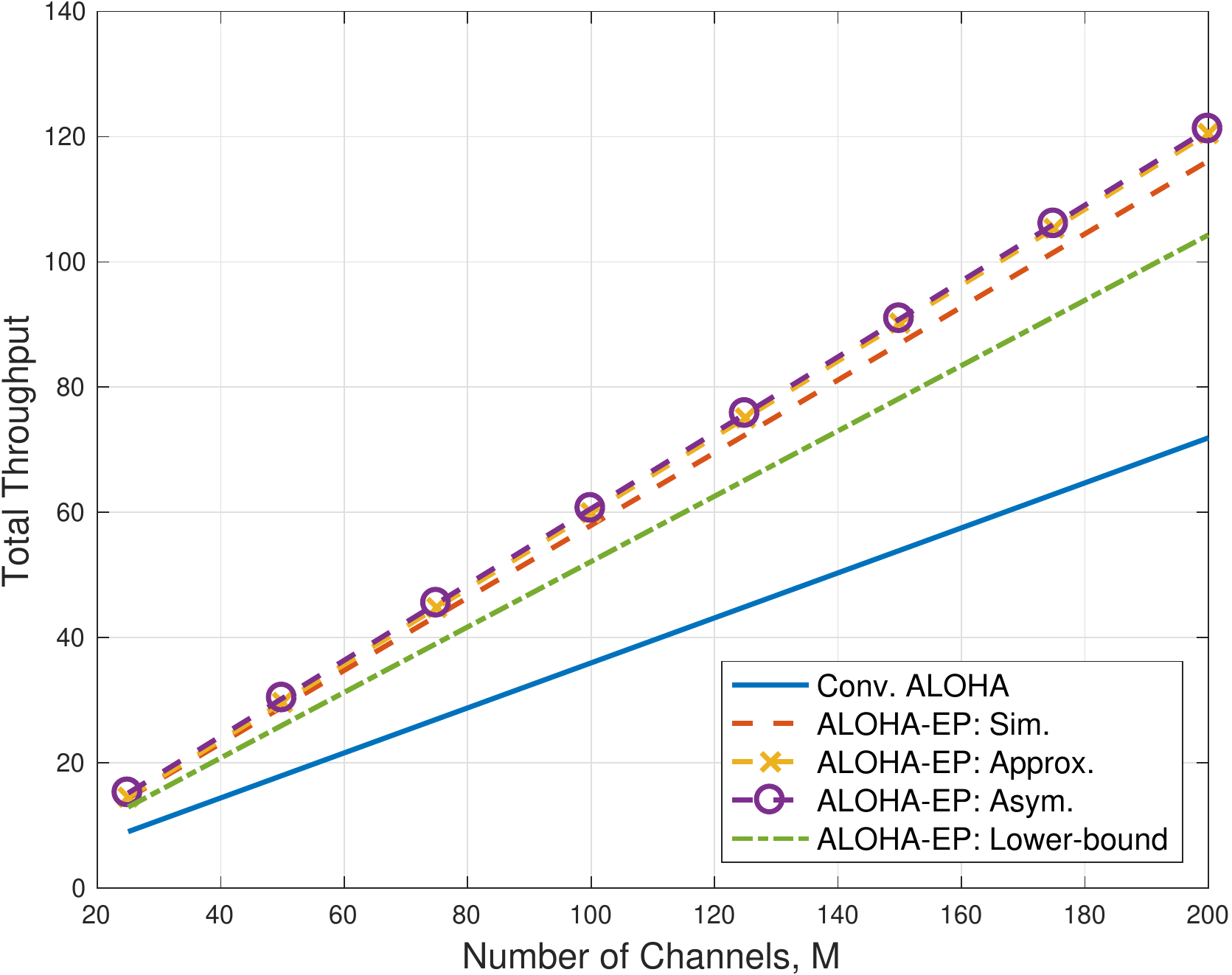}
\end{center}
\caption{Total throughputs 
of conventional multichannel ALOHA and
multichannel ALOHA with EP as functions
of $M$ when $\alpha = \frac{\lambda}{M} = 0.8$.}
        \label{Fig:Tplt3}
\end{figure}

In Fig.~\ref{Fig:Tplt3}, the total throughputs 
of conventional multichannel ALOHA and
multichannel ALOHA with EP are shown as functions
of $M$ when $\alpha = \frac{\lambda}{M} = 0.8$ is fixed. 
As shown by the lower-bound in \eqref{EQ:Mgap},
the difference between the throughputs grows
linearly with $M$. 

\begin{figure}[thb]
\begin{center}
\includegraphics[width=\figwidth]{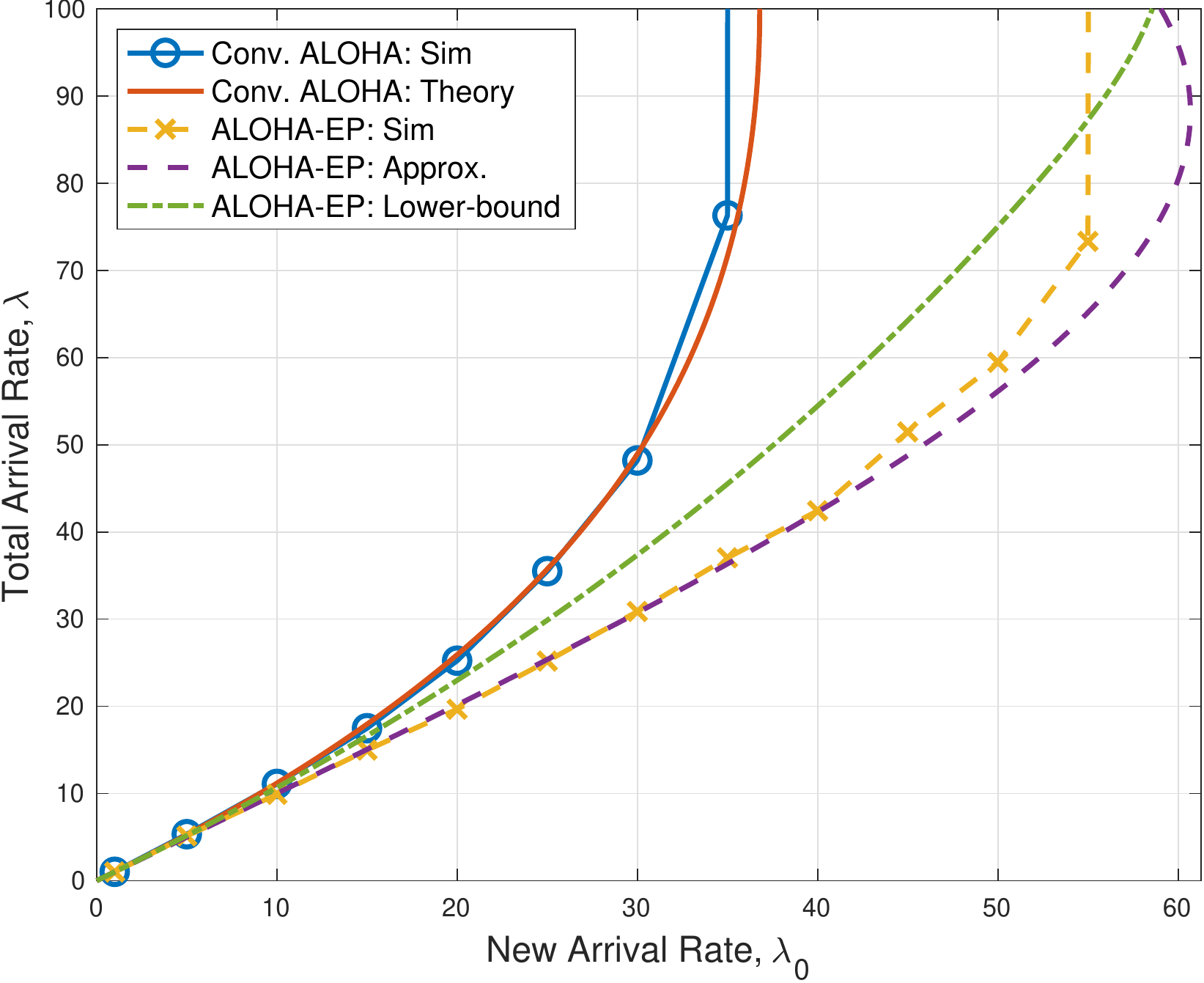} \\
(a) \\
\includegraphics[width=\figwidth]{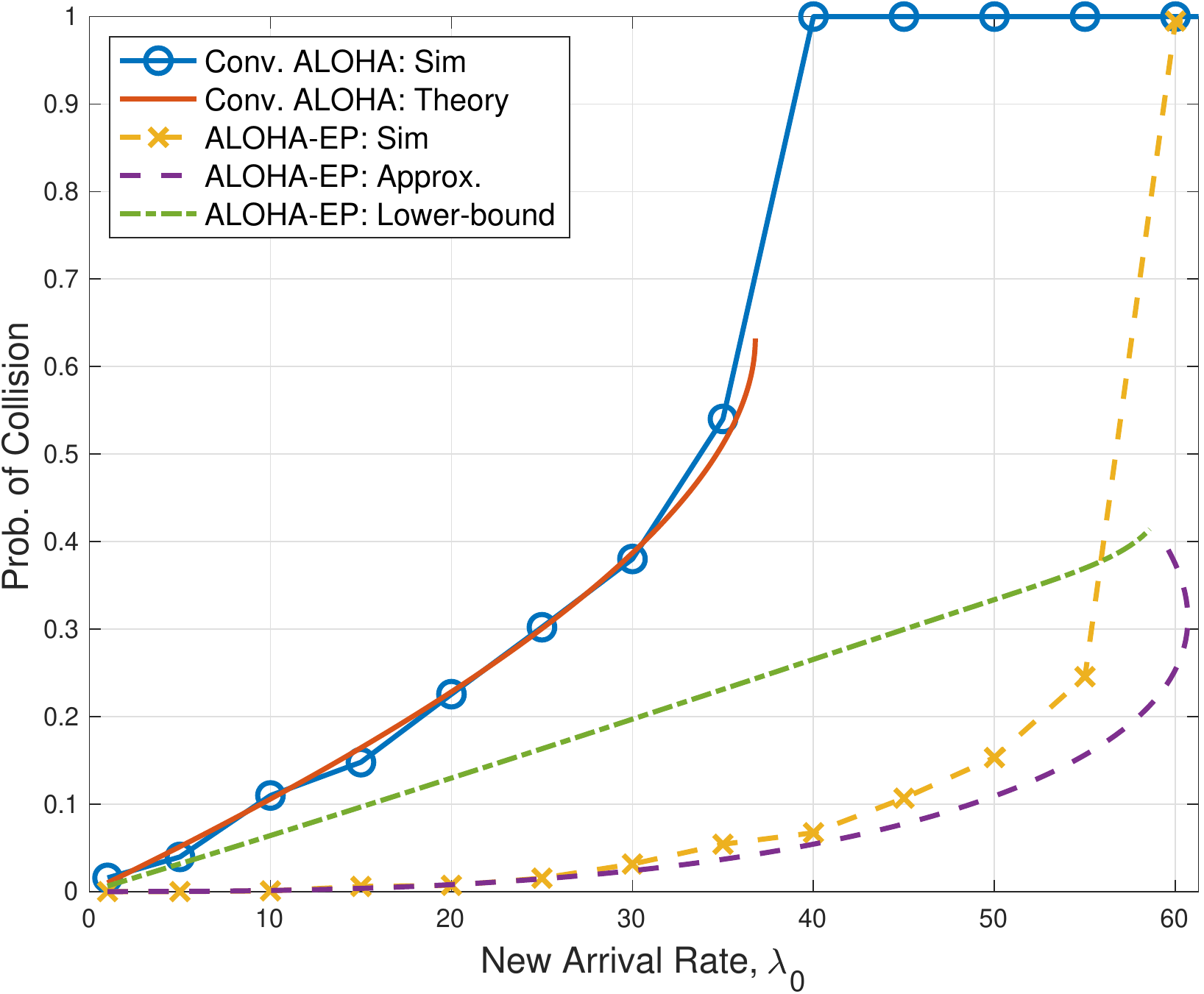} \\
(b) \\
\end{center}
\caption{Performance of conventional multichannel ALOHA
and multichannel ALOHA with EP when fast retrial is used with $M = 100$:
(a) the relationship between $\lambda$ and $\lambda_0$;
(b) the relationship between the probability of collision and $\lambda_0$.}
        \label{Fig:Splt1}
\end{figure}

We now consider the simulation results with fast retrial,
where 
collided packets can be re-transmitted in the next slot.
In Fig.~\ref{Fig:Splt1} (a),
the relationship between $\lambda$ and $\lambda_0$
is shown for conventional multichannel ALOHA 
and multichannel ALOHA  with EP. It can be observed
that the theoretical relationship between
$\lambda$ and $\lambda_0$ under the assumption
that the total of the new and back-logged arrivals
follows a Poisson distribution agrees with
simulation results when $\lambda_0$ is sufficiently
low
($\lambda_0 \le 30$ for conventional multichannel ALOHA and
$\lambda_0 \le 40$ for multichannel ALOHA with EP).
We also observe that multichannel ALOHA with EP
can have a smaller number of back-logged packets
than conventional multichannel ALOHA, since $\lambda$ 
of multichannel ALOHA with EP is smaller than 
that of conventional multichannel ALOHA
at the same value of $\lambda_0$.
This is due to the fact that the probability of collision
of 
multichannel ALOHA with EP is lower than 
that of 
conventional multichannel ALOHA as shown in Fig.~\ref{Fig:Splt1} (b).

Note that the steady-state analysis in 
Section~\ref{S:SS}
is based on the assumption that $\lambda$ follows
a Poisson distribution as in \eqref{EQ:lamlam}.
In general, if there are a number of collided packets,
this assumption cannot be used \cite{YJChoi06}.
Furthermore, the approximation in \eqref{EQ:L2} and
the asymptotic expression in \eqref{EQ:psi}
are reasonable when $\lambda < M$.
As a result, the analysis in Section~\ref{S:SS} can be used
only when $\lambda_0$ is sufficiently lower than $M$
or for a lightly loaded system, which is in fact the case that
we are interested in to consider the exploration for multichannel ALOHA
as mentioned earlier.

\section{Concluding Remarks}
\label{S:Conc}

In this paper, an exploration 
approach has been proposed for multichannel ALOHA by sending preambles
prior to packet transmissions to allow
active users to learn the state of contention.
We found that the exploration gain is 
$2 - e^{-1}$ in terms of the ratio 
of the maximum throughput
of multichannel ALOHA with EP to that of 
conventional multichannel ALOHA.
As a result, with the same bandwidth,
compared with conventional multichannel ALOHA,
we could see that the proposed multichannel ALOHA
with EP can support more sensors and devices in MTC.
We also showed that the exploration
can significantly decrease the delay outage probability
when the system with fast retrial is lightly loaded.

\bibliographystyle{ieeetr}
\bibliography{mtc}

\end{document}